\documentclass[cits]{PoS}

\usepackage{textcomp}
\usepackage[latin1]{inputenc}
\usepackage{amsfonts}
\usepackage{amssymb}
\usepackage{amsbsy}
\usepackage{amsmath}
\usepackage{xspace}
\usepackage{graphicx}

\newcommand{\invfb}{\,{\rm fb}^{-1}}

\newcommand{\gev}{{\hbox{GeV}}}
\newcommand{\mmev}{{\hbox{MeV}/c^2}}

\newcommand{\mbc}{{M_{\textrm{bc}}}}
\newcommand{\deltae}{{\Delta E}}

\newcommand{\lint}{{L_{\textrm{int}}}}

\newcommand{\BR}{{\mathcal B}}

\newcommand{\FourS}{\Upsilon(4S)}
\newcommand{\FiveS}{\Upsilon(5S)}
\newcommand{\bs}{B_s^0}
\newcommand{\barbs}{{\bar B}_s^0}
\newcommand{\bsst}{{B_s^{\ast}}}
\newcommand{\barbsst}{{{\bar B_s}^{\ast}}}
\newcommand{\bsST}{{B_s^{(\ast)}}}

\newcommand{\KS}{{K_S^0}}

\newcommand{\jpsi}{{J\!\!/\!\!\psi}}
\newcommand{\fz}{f_0}

\newcommand{\bsSTbsST}{{\bsST{\bar B_s}^{(\ast)}}}

\newcommand{\bsdspi}{{\bs\to D_s^-\pi^+}}

\newcommand{\bsjpsifz}{{\bs\to\jpsi\,\fz}}
\newcommand{\bsjpsifzu}{{\bs\to\jpsi\,\fz(980)}}
\newcommand{\bsjpsifzd}{{\bs\to\jpsi\,\fz(1370)}}
\newcommand{\bsLcpiL}{{\barbs\to\Lambda_c^+\pi^-\bar\Lambda}}

\newcommand{\fss}{{\left(90.1^{+3.8}_{-4.0}\pm0.2\right)\%}}

\title{Recent Belle results from $\FiveS$ sample}

\ShortTitle{Recent Belle results from $\FiveS$ sample}

\author{\speaker{Remi Louvot}\\
  (On behalf of the Belle collaboration)\\
  Laboratoire de Physique des Hautes \'Energies,\\
  \'Ecole Polytechnique F\'ed\'erale de Lausanne~(EPFL), Lausanne, Switzerland\\
  E-mail: \email{remi.louvot@epfl.ch}}

\abstract{

  The large data sample recorded with the Belle detector at the
  $\FiveS$ energy provides a unique opportunity to study the poorly-known $\bs$ meson.
  Two analyses, performed with a data sample representing an integrated luminosity of 121~$\invfb$, are presented: the measurement of the $\bsjpsifzu$ and $\bsjpsifzd$ branching fractions, and the $5\sigma$ observation of the decay $\bsLcpiL$ which is the first observation of a baryonic $\bs$ decay.
  In addition, we present new results of a measurement of the CKM angle $\phi_1(\beta)$ with $B\pi$ tagged events.

 ~

~

29 October 2011

LPHE Note 2011-04}

\FullConference{XXIst International Europhysics Conference on High Energy Physics\\
		 21--27 July 2011\\
		 Grenoble, Rh\^one-Alpes, France}

\begin{document}

\section*{Introduction}

The Belle experiment \cite{NIMA_479_117}, located at the interaction point of
the KEKB asymmetric-energy $e^+e^-$ collider,
was designed for the study of $B$ mesons\footnote{The notation ``$B$'' refers either to a $B^0$ or a $B^+$.
  Moreover, charge-conjugated states are implied everywhere.}
produced in $e^+e^-$ annihilation at a center-of-mass (CM) energy corresponding to the mass of
the $\FourS$ resonance ($\sqrt s\approx10.58~\gev$).
However, a data sample of integrated luminosity $\lint=121\invfb$ has been recorded and analyzed at the energy of the $\FiveS$ resonance
($\sqrt s\approx10.87~\gev$), above the $\bs\barbs$ threshold.

Apart from the $e^+e^-\to u\bar u, d\bar d, s\bar s, c\bar c$ continuum events,
the $e^+e^-\to b\bar b$ process can produce different kinds of final states involving
a pair of non-strange $B$ mesons \cite{PRD_81_112003} ($B^{\ast}\bar B^{\ast}$,
$B^{\ast}\bar B$, $B\bar B$, $B^{\ast}\bar B^{\ast}\pi$, $B^{\ast}\bar B\pi$,
$B\bar B\pi$, $B\bar B\pi\pi$ and $B\bar B\gamma$),
a pair of $\bs$ mesons ($\bsst\barbsst$, $\bsst\barbs$ and $\bs\barbs$),
or final states involving a light bottomonium resonance below the open-beauty
threshold \cite{PRL_100_112001}.
The $B^{\ast}$ and $\bsst$ mesons always decay by emission of a photon.
The total $e^+e^-\to b\bar b$ cross section at the $\FiveS$ energy was measured
to be $\sigma_{b\bar b}=302\pm14$~pb
\cite{PRL_98_052001} and the fraction of $\bs$ events to be
$f_s=\sigma(e^+e^-\to\bsSTbsST)/\sigma_{b\bar b}=(19.3\pm2.9)$\% \cite{PDG10}.
The dominant $\bs$ production mode, $b\bar b\to\bsst\barbsst$,
represents $f_{\bsst\barbsst}=\fss$ of the $b\bar b\to\bsSTbsST$ events, as measured with $\bsdspi$ events \cite{PRL_102_021801}.

$\bs$ candidates are fully reconstructed
from the final-state particles.
From the reconstructed four-momentum in the $e^+e^-$ center-of-mass, $(E_{\bs}^{\ast},\pmb{p}_{\bs}^{\ast})$,
two observables are used to extract the signal yield:
the energy difference $\deltae=E_{\bs}^{\ast}-\sqrt s/2$ and the
beam-constrained mass $\mbc=\sqrt{s/4-\pmb{p}_{\bs}^{\ast2}}$.
The corresponding branching fraction is then computed using the total efficiency (including sub-decay branching fractions)
determined with Monte-Carlo (MC) simulations, $\sum\varepsilon\BR$,
and the number of $\bs$ mesons produced via the $e^+e^-\to\bsst\barbsst$ process, $N_{\bs}=2\times\lint\times\sigma_{b\bar b}\times f_s\times f_{\bsst\barbsst}$.

\section{Study of $\pmb{\barbs\to\Lambda_c^+\pi^-\bar\Lambda}$}\label{sec:lambda}

The $\bsLcpiL$ decay is the counterpart if the already-observed $B^-\to\Lambda_c^+\pi^-\bar p$ decay. The study of $B_{(s)}$ baryonic decays is important as the latest observations \cite{masspeak} exhibit a baryon-antibaryon mass peak near the kinematic threshold and tend to have larger branching fractions than two-body decays.

We fully reconstruct the decay via $\Lambda_c^+\to pK^-\pi^+$ and $\bar\Lambda\to\bar p\pi^+$. After a fit of the two $\Lambda_{(c)}$ vertices, only $\barbs$ candidates for which the  $\Lambda_c^+$ ($\bar\Lambda$) invariant mass lies within 100 $\mmev$ (4 $\mmev$) of the PDG value \cite{PDG10} are retained. The continuum is rejected with requirements on second-to-zeroth Fox-Wolfram moment ratio \cite{FoxWolfram}, $R_2<0.5$, and the cosine of thrust angle, $\cos\theta_{\rm th}<0.85$.

\begin{figure}
\centering

\includegraphics[width=0.9\linewidth,height=6cm]{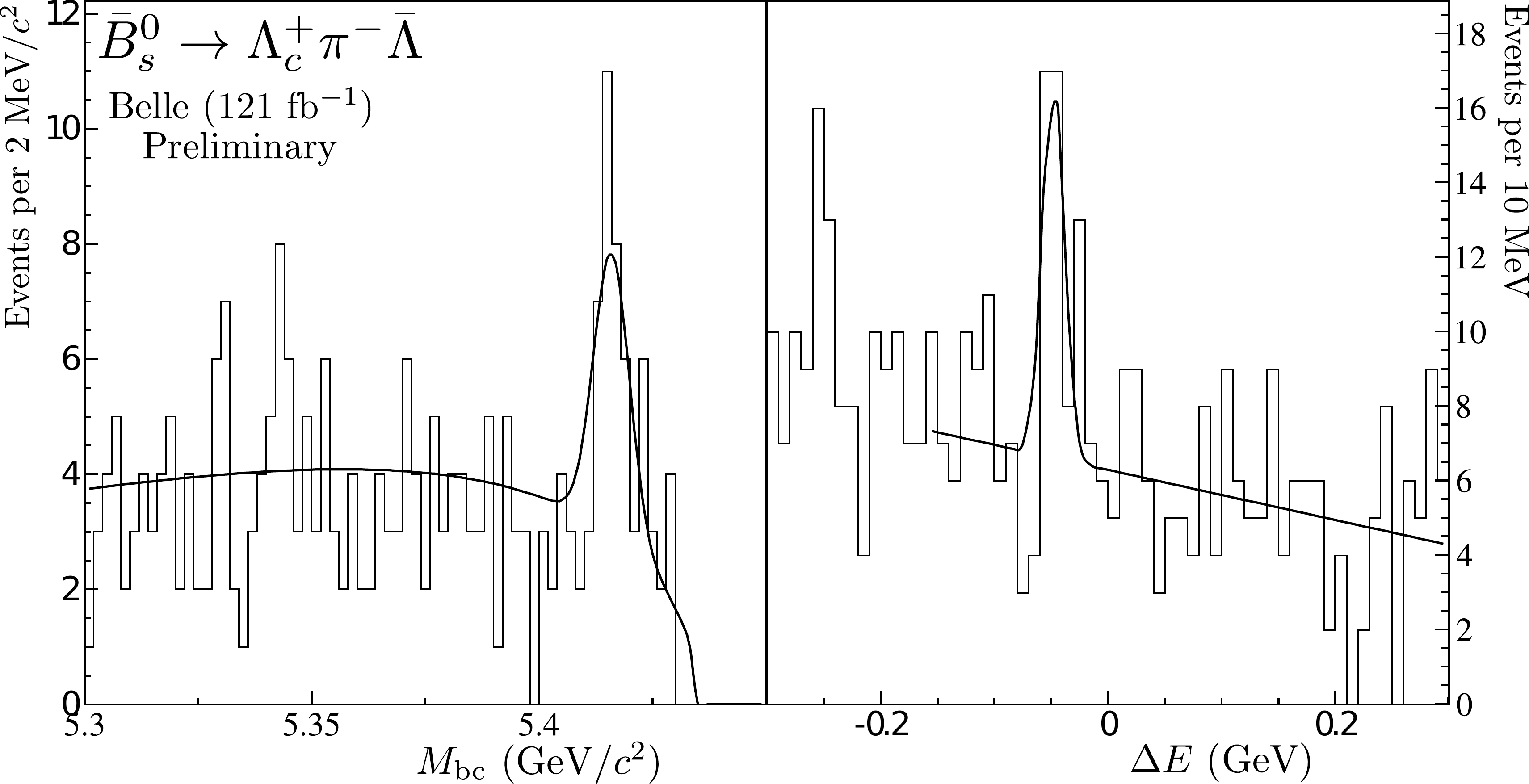}
\caption{\label{lfig}$\mbc$ (left) and $\deltae$ (right) distributions of the $\bsLcpiL$ candidates (histogram) together with the fit result (solid curve). The dotted curve shows its background component.}
\end{figure}

A two-dimensional binned fit on $\mbc$ and $\deltae$ leads to a first $5.0\sigma$-significant (including systematic effects) observation of $24\pm7$ events (Fig.~\ref{lfig}). This is the first observation of a $\bs$ baryonic decay. The measured branching fraction, $$\BR(\bsLcpiL)=(4.8\pm1.4({\rm stat.})\pm0.9({\rm syst.})\pm1.3(\Lambda_c^+))\times10^{-4}\,,$$ where the uncertainty due to the $\Lambda_c^+$ branching fraction is quoted separately, is compatible with that of $B^-\to\Lambda_c^+\pi^-\bar p$ \cite{PDG10}.

\newpage

\section{Study of $\pmb{\bs\to J\!/}\!\mathbf{\psi}\,\pmb{f_0}$}\label{sec:jpsi}
$\bs$ decays to $CP$ eigenstates are important for $CP$-violation measurements \cite{PRD_63_114015}.
The $\bsjpsifz$ mode is especially interesting for the hadron-collider experiments because it can be reconstructed from charged tracks only.

The $\jpsi$ candidates are formed with oppositely-charged electron or muon pairs,
while $f_0$ candidates are formed with $\pi^+\pi^-$ pairs. 
A mass and vertex constrained fit is then applied to the $\jpsi$ candidates.
If more than one candidate per event satisfies all the selection criteria, the one with the $\mbc$ value the closest to the expected signal mean is selected.
The main background is the continuum, which is reduced by requiring $R_2<0.4$. 
The $\bsjpsifz$ signal is fitted using the energy difference, $\deltae$, and the $\fz$ mass, $M_{\pi^+\pi^-}$, distributions. Two $f_0$ resonances, $f_0(980)$ and $f_0(1370)$, are included in the fit.

We obtain a 8.4$\sigma$ observation of $63^{+16}_{-10}$ $\bsjpsifzu$ events and the first evidence for $\bsjpsifzd$ with $19^{+6}_{-8}$  events \cite{PRL_106_121802}. We extract the branching fractions $\BR(\bs\to\jpsi\fz(980);$ $\fz(980)\to\pi^+\pi^-)=[1.16^{+0.31}_{-0.19}({\rm stat.})^{+0.15}_{-0.17}({\rm syst.})^{+0.26}_{-0.18}(N(\bs))]\times 10^{-4}$
and
  $\BR(\bs\to\jpsi\fz(1370);$ $\fz(1370)\to\pi^+\pi^-)=[0.34^{+0.11}_{-0.14}({\rm stat.})^{+0.03}_{-0.02}({\rm syst.})^{+0.08}_{-0.05}(N(\bs))]\times10^{-4}$, which are in agreement with other ha\-dron-collider experiments \cite{PLB_698_115}.

\section{Measurement of $\pmb{\sin 2\phi_1}$ with $\pmb{B\pi}$ tagging}

Because the $\FiveS$ mass is above the $B^{\ast}\bar B^{\ast}\pi$ threshold, a significant number of $\FiveS\to B^{(\ast)}\bar B^{(\ast)}\pi^{\pm}$ events are present in the data sample \cite{PRD_81_112003}. The sign of the pion indicates whether the event contains a $B^{(\ast)0}$ ($e^+e^-\to B^{(\ast)0}B^{(\ast)-}\pi^+$) or a $\bar B^{(\ast)0}$ ($e^+e^-\to\bar B^{(\ast)0}B^{(\ast)+}\pi^-$). With $B^0$ decaying to a $CP$ eigenstate, the asymmetry, $A_{BB\pi}=(N(BB\pi^-)-N(BB\pi^+))/(N(BB\pi^-)+N(BB\pi^+))$, the CKM angle $\phi_1$ can be determined via the relation \cite{NPB_405_55}: $\sin2\phi_1=-\eta_{CP}A_{BB\pi}(1+x^2)/x$, where $x=\Delta m/\Gamma$.

From a clean sample of $75.9\pm9.5$ fully reconstructed $B^0\to\jpsi(\to l^+l^-)K_S^0(\to\pi^+\pi^-)$ events, we simultaneously fit the missing masses of the $B^0\pi^-$ and $B^0\pi^+$ candidates by adding a charged pion.
The fit involves three signal components for the $B^{\ast}\bar B^{\ast}\pi$, $B^{\ast}\bar B\pi$ (+c.c.) and $B\bar B\pi$ classes of events.
A total signal of $21.5\pm6.8$ $B^0\pi^{\pm}$ events is obtained together with the asymmetry $A_{BB\pi}=0.28\pm0.28$(stat.). While this analysis clearly suffers from lack of statistics, it nevertheless demonstrates that $\phi_1$ can be measured by this alternative method.

\begin{figure}[t]
  \centering
  \includegraphics[width=0.9\linewidth]{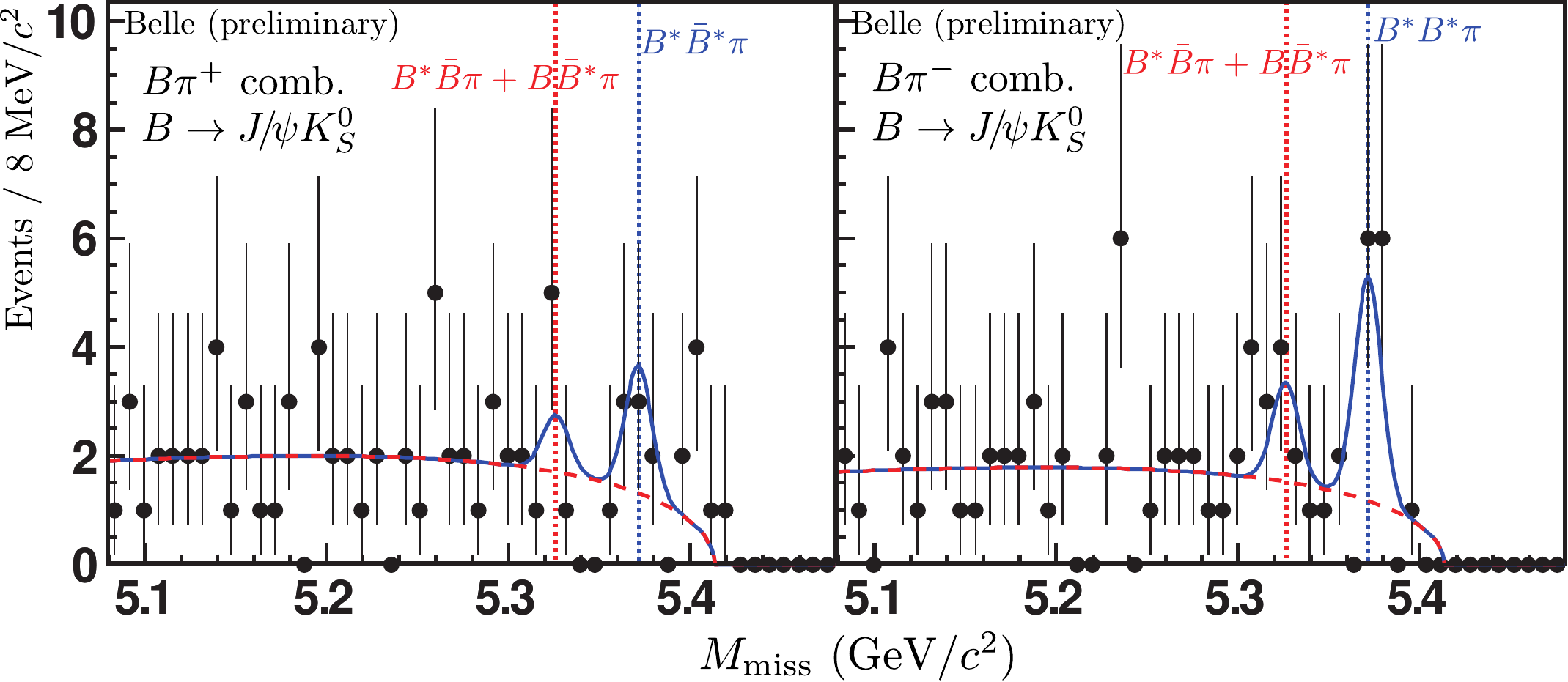}
  \caption{\label{fig:sato}$B^0\pi^+$ (left) and $B^0\pi^-$ (right) missing mass distributions for selected $B^0\to\jpsi\KS$ candidates (data points) together with the fit result (solid curve) and its background component (dashed curve).}
\end{figure}

\section*{Conclusion}
We presented new results on $\bs$ decays obtained from 121 $\invfb$ of $\FiveS$ data recorded by the Belle detector.
While modes with large statistics can provide precise measurements of branching fractions and $\bsST$ properties, 
first observations of several $CP$-eigenstate $\bs$ decays are a confirmation
of the large potential of our 120$\invfb$ $e^+e^-\to\FiveS$ data sample and advocate an ambitious $\bs$ program at super-$B$ factories.

\end{document}